\documentclass[10pt, conference, compsocconf]{IEEEtran}
\usepackage{times}
\usepackage{url}
\usepackage{graphics}
\usepackage{color}
\usepackage{multicol,multienum}

\usepackage{dcolumn}
\newcolumntype{d}[1]{D{.}{.}{#1} }

\usepackage{arydshln}

\usepackage{graphicx}
\usepackage{subfigure}
\usepackage{amssymb}
\usepackage{latexsym}
\usepackage{algorithmic}
\usepackage{amsmath}
\usepackage{multirow}
\usepackage{bigstrut}
\usepackage{array}
\usepackage{tabularx}
\usepackage{graphicx}

\newtheorem{example}{Example}

\graphicspath{{figures/}}

\begin{document}
%
%
\title{Extended Report: Fine-grained Recognition of Abnormal Behaviors for Early Detection of Mild Cognitive Impairment}



\author{
\IEEEauthorblockN{Daniele Riboni$^{\ast}$ \hspace{0.2cm} Claudio Bettini$^{\ast}$ 
\hspace{0.2cm} Gabriele Civitarese$^{\ast}$ \hspace{0.2cm} Zaffar Haider Janjua$^{\ast}$ 
\hspace{0.2cm} Rim Helaoui$^{\diamond}$}
\IEEEauthorblockA{
$^{\ast}$Universit\`a degli Studi di Milano, Dept. of Computer Science, EveryWare Lab\\
$^{\diamond}$University of Mannheim, Research Group Data \& Web Science\\
Email: \{daniele.riboni, claudio.bettini, gabriele.civitarese, zaffar.janjua\}@unimi.it, 
rim@informatik.uni-mannheim.de}}

\maketitle

\begin{abstract}
According to the World Health Organization, the rate of people aged 60 or more is growing faster than any other age group in almost every country, and this trend is not going to change in a near future. Since senior citizens are at high risk of non communicable diseases requiring long-term care, this trend will challenge the sustainability of the entire health system. Pervasive computing can provide innovative methods and tools for early detecting the onset of health issues. In this paper we propose a novel method to detect abnormal behaviors of elderly people living at home. The method relies on medical models, provided by cognitive neuroscience researchers, describing abnormal activity routines that may indicate the onset of early symptoms of mild cognitive impairment. A non-intrusive sensor-based infrastructure acquires low-level data about the interaction of the individual with home appliances and furniture, as well as data from environmental sensors. Based on those data, a novel hybrid statistical-symbolical technique is used to detect the abnormal behaviors of the patient, which are communicated to the medical center. Differently from related works, our method can detect abnormal behaviors at a fine-grained level, thus providing an important tool to support the medical diagnosis. In order to evaluate our method we have developed a prototype of the system and acquired a large dataset of abnormal behaviors carried out in an instrumented smart home. Experimental results show that our technique is able to detect most anomalies while generating a small number of false positives.
\end{abstract}


\section{Introduction}
\label{sec:introduction}
%
Several recent studies show that the proportion of elderly people over the whole population is rapidly growing in most countries. For instance, the European old-age dependency ratio (i.e., the ratio of people aged 65 years or older to people aged 15-64 years) is projected to double in the next decades~\cite{lancet13}. As a consequence, a growing portion of people is at high risk of experiencing non communicable diseases, frailty and social exclusion, and may need long-term care, including nursing at home or frequent hospitalization. Of course, the inability of living independently may not only spoil the quality of life of elderly people and of those caring for them, but will also challenge the sustainability of the entire health system. Hence, there is a growing interest in exploiting pervasive computing technologies to support independent living and healthcare, especially for the senior population.

In this paper we propose a novel method to support early detection of mild cognitive impairment (MCI) for elderly people living independently at home. In the medical literature, MCI is used as a clinical diagnosis to describe a transitional state between healthy cognitive ageing and dementia, characterized by preserved functional abilities~\cite{PSW99}. According to the criteria proposed by the International Working Group on MCI, there are evidences of subtle differences in performing instrumental activities of daily living (IADLs) among MCI patients compared to both healthy older adults and individuals with dementia~\cite{WPK04}. Hence, long-term monitoring of daily living activities and recognition of abnormal behaviors may help practitioners to early detect the onset of cognitive impairment. 

Different scales have been proposed to assess the cognitive health of people based on questionnaires and interviews about the ability of performing various kinds of IADLs~\cite{scales}. However, that approach is prone to reporting bias; moreover, it cannot be applied to continuously monitor the cognitive health of a large number of people, since it incurs evident overheads in terms of time, resources and monetary costs. A few previous works, reviewed in Section~\ref{sec:related}, have tried to detect behavioral markers of MCI onset through pervasive computing technologies, obtaining significant correlation between the predicted and actual cognitive status of the patient. However, those works have different limitations. Some of them require the execution of ability tests about the performance of IADLs in an instrumented smart home of a hospital; hence, they incur high costs and cannot be applied on a continuous basis. Other works rely on continuous monitoring of low-level behavioral markers (steps taken, walking speed\ldots). While potentially useful to trigger alarms about possible MCI onset, those markers do not provide specific support to the diagnosis, since they do not report fine-grained descriptions of the anomalies occurred during the execution of IADLs.

In order to overcome the limitations of existing techniques we propose FABER, a novel technique for Fine-grained Abnormal BEhavior Recognition. Our method relies on medical models describing abnormal activity routines that may indicate the onset of early symptoms of MCI. These models have been acquired through the collaboration with cognitive neuroscience experts of a leading center for care and research on neurodegenerative disorders. A non-intrusive sensor-based infrastructure is deployed at the patient's home, which acquires low-level information about the interaction with home appliances and furniture, as well as environmental parameters. Based on sensor data, we first detect the general activity being performed by the subject and then recognize \emph{anomalies} in performing that activity or a group of activities. They include inappropriate timing and unnecessary repetitions of subactions, but also high level observations like ``irregularly assuming meals" or ``often consuming cold meals". We use a hybrid statistical-symbolical technique including supervised learning, rule-based reasoning and probabilistic reasoning. Abnormal behaviors are communicated to the medical center for further analysis and interpretation.  
Differently from previous works, our method can be applied continuously at the patient's home and, thanks to symbolic reasoning postprocessing over recognized activities, abnormal behaviors can be detected at a fine-grained level.  In order to evaluate our approach, we have developed a prototype of our system, and collected a large dataset from an instrumented smart home. Experimental results show that our technique is able to detect most of the abnormal behaviors that we have targeted while producing a small number of false positives.

This manuscript is an extended report of~\cite{PerCom15}.
The rest of the paper is structured as follows. 
Section~\ref{sec:related} discusses related work. 
Section~\ref{sec:technique} presents the FABER method. 
Section~\ref{sec:experiments} reports experimental results. 
Finally, Section~\ref{sec:conclusions} concludes the paper.




\section{Related work}
\label{sec:related}
Several studies in the neuropsychology research field show that it is possible to distinguish between cognitively healthy adults and cognitively impaired individuals based on subtle differences in their behavioral patterns~\cite{WPK04}. There is a growing interest in exploiting pervasive computing technologies to automatically capture and measure those differences~\cite{PS13}. For instance, a sensor-based infrastructure has been used to unobtrusively monitor the execution of IADLs by older adults in a smart-home~\cite{SSC13}; the results have shown a significant correlation between the cognitive health status of the subject and the level of assistance that he needed in order to complete the activities. More recently, motion sensors and contact sensors have been used in~\cite{HAA08} to measure low-level activity patterns, such as  walking speed and activity level in the home; results have shown that the coefficient of variation in the median walking speed is a statistically significant measure to distinguish MCI subjects from healthy seniors.

Based on this line of research, a few works have proposed to apply artificial intelligence methods on data acquired in sensor-rich environments, for assessing the cognitive health status of an individual performing a fixed set of predefined activities. In the work of Dawadi et al.~\cite{DCS13}, patients were invited to execute a list of routines (e.g., write a letter, prepare lunch) inside a hospital smart-home. Different kinds of sensors were used to detect movements inside the home and to track the use of furniture and appliances. Based on data coming from the home sensors, machine learning methods were used to assign a score to each performed activity; the score measures the ability of the subject to perform the activity correctly. The achieved scores were then used to predict the cognitive status of the patient (cognitive health or dementia). However, experimental results showed a not completely satisfactory degree of correlation among the predicted scores and the ones assigned by a human observer. In general, the low correlation may be due to the intrinsic difficulty of capturing the variability of human behaviors from a corpus of training data. In this work, we take a different approach: we use supervised learning only to detect the start- and end-times of activities, while we rely on domain knowledge provided by neuroscience experts to recognize the actual anomalies. To the best of our knowledge, our work is the first one that tries to apply this approach to cognitive health assessment.

The supervised learning approach has been applied in other works, including~\cite{DCS13b,DCP11,CS09}, using several other learning methods. A further difference  with those works is that they assume that the patient executes a predefined set of IADLs following the instructions of practitioners in a medical center, while our method is intended to run continuously at the patient's home, and does not interfere with the normal behavior of the patient.

Finally, we mention that several research efforts have been made to automatically detect abnormal behaviors for surveillance applications. Typically, in that field, abnormal actions are recognized based on the analysis of audio and video and on the application of machine learning techniques~\cite{OW12}. However, audio- and video-based systems are not suitable to our problem due to obvious privacy issues. Moreover, surveillance systems are mainly targeted to low-level physical actions, such as assaults, fights, stealing of objects, while our goal is to monitor high-level daily living behaviors, which are subject to high variability of execution based on the characteristics of the specific individual, on the environment and on many other contextual conditions.

\section{The FABER hybrid technique}
\label{sec:technique}
In this section we illustrate the \emph{Fine-grained Abnormal 
BEhavior Recognition (FABER)} hybrid technique to support
early detection of MCI.

\subsection{Recognition framework}
\label{subsec:framework}
\begin{figure}[t!]
  \centering
     \includegraphics[width=\columnwidth]{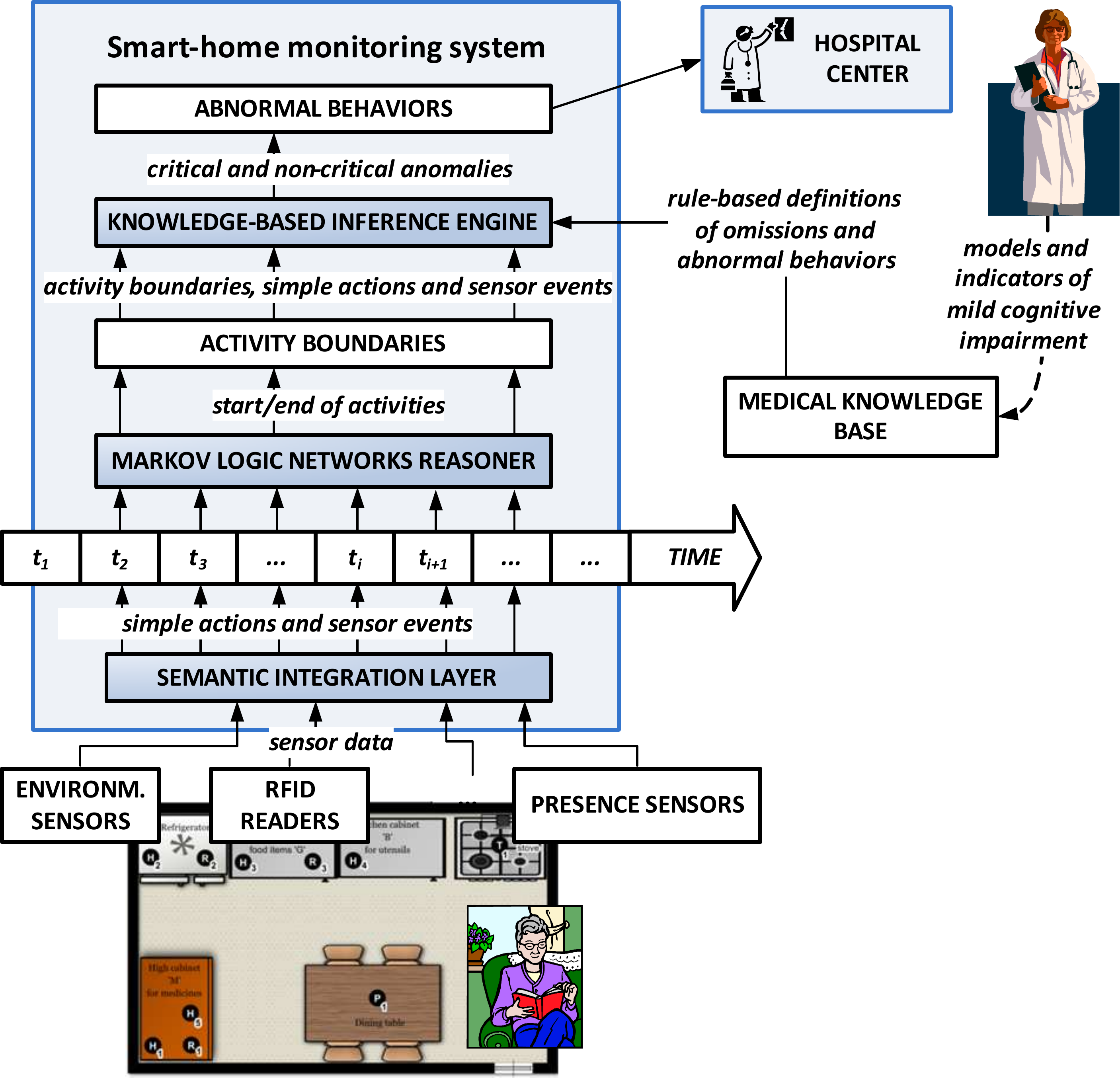}
  \caption{Recognition framework}
  \label{fig:arch}
\end{figure}

%
In Figure~\ref{fig:arch} we show the recognition framework. 
The system is implemented at the elderly's 
home. A smart-home monitoring system running on a mobile device
(e.g., a tablet) is in charge of executing the
FABER algorithms. 
Different sensors, including environmental sensors, presence
sensors, and RFID readers, are attached to furniture and 
instruments, and communicate raw data to the 
\textsc{semantic integration layer}. 
That layer is in charge of using raw sensor data to 
detect simple actions (e.g., ``the fridge door has been opened'')
and other events (e.g., ``the temperature in the kitchen
is more than 30 degrees Celsius''). Actions and events, together
with their timestamps, are communicated to the 
\textsc{markov logic network (mln) reasoner} using a shared vocabulary.
The reasoner periodically (e.g., daily) analyzes the event logs and 
infers the start/end time of activities based on the received data.
The inferred activity boundaries are communicated --together with 
actions and events-- to the \textsc{knowledge-based inference engine}. 
The inference engine evaluates the rules modeling abnormal behaviors, 
which are extracted from a medical knowledge base of MCI models and 
indicators.
Finally, detected abnormal behaviors are communicated to the hospital
center for further analysis by the doctors.

\subsection{Semantic integration of sensor data}
\label{subsec:semint}
%
The \textsc{semantic integration layer} is in charge of acquiring
raw sensor data and to use them for inferring semantic
descriptions of the current context, which are exploited by the 
MLN reasoner to detect the activity boundaries.
Depending on the kind of available sensors, that module applies
simple inference methods to derive basic actions and events. 
For instance, a rule states that ``if the presence sensor detects 
a presence near the kitchen table, and the sensor on the kitchen's 
chair detects a weight higher than $50$Kg, then the current action 
is \emph{sitting at the kitchen chair}''.
Timestamped actions and events are represented using a shared vocabulary
and communicated to the MLN reasoner.

\subsection{Detection of activity boundaries}
After presenting our temporal model, we illustrate Markov Logic Networks (MLN)
and we explain how we use this probabilistic logic to detect activity boundaries. 

\subsubsection{Temporal model}
Suppose that, in our system, the following temporal sequence of sensor events occurs: 
\[
\langle \, event(e_{j_1}, t_1), event(e_{j_2}, t_2), \ldots, event(e_{j_m}, t_m) \, \rangle,
\]
where $event(e_{j_i},t_i)$ indicates that the sensor event $e_{j_i}$ occurred at time instant $t_i$. 
For the sake of this work, we assume that sensor nodes communicate their sensed events in 
real-time to a gateway. The gateway is in charge of assigning a unique timestamp to each event, 
based on the time at which it is received. 
Hence, we impose a total order on event timestamps $\langle \, t_1, t_2, \ldots, t_m \, \rangle$.

\subsubsection{MLN}
\label{subsec:mln}
Thanks to its rich expressiveness, first-order logic (FOL) and some of its fragments have been used in different works to model and reason with human activities~\cite{RiboniPMC11}. However, it has been recognized that FOL knowledge bases are too inflexible to model many real-world scenarios. This limitation is even more accentuated in applications involving the temporal context. In particular, in the domain of activity recognition, temporal sequences of sensor events can be ambiguous to interpret. Indeed, the same sensor event sequence can result from the execution of different and possibly conflicting activities. 

\begin{example}
\label{ex:act}
Consider the following first-order logic (FOL) knowledge base:
\begin{eqnarray}
\forall \, a, \, e_{j}, \, e_{k}, \, t_i, \, t_{i+1} \,\, event(e_{j}, t_i) \wedge event(e_{k}, t_{i+1}) 
\\\nonumber \rightarrow currentActivity(a, t_i).\\
\forall \, t \,\, currentActivity(SetTheTable, t) 
\\\nonumber \rightarrow \neg \, currentActivity(WashDishes, t).
\end{eqnarray}
Formula~(1) states that the temporal sequence of two sensor events $e_j$ and $e_k$ occurring at $t_i$
and $t_{i+1}$, respectively, indicates the execution of an activity $a$ at $t_i$. 
Formula~(2) states that the current activity of an individual cannot be ``set the table''
and ``wash dishes'' at the same time instant.
Suppose to instantiate event $e_{j}$ to \emph{ClosingSilverwareDrawer} and $e_k$ to 
\emph{OpeningGlasswareCabinet}. 
The rules below, obtained by grounding formula (1), encode the fact that the occurrence 
of the temporal sequence $\overline{s} = \langle$ \emph{event(ClosingSilverwareDrawer, $t_i$)}, 
\emph{event(OpeningGlasswareCabinet, $t_{i+1}$)} $\rangle$ can indicate both activities 
``set the table'' and ``wash dishes'' at $t_i$:
\begin{eqnarray}
event(ClosingSilverwareDrawer, t_i) \wedge 
\\\nonumber event(OpeningGlasswareCabinet, t_{i+1}) 
\\\nonumber \rightarrow currentActivity(SetTheTable, t_i).\\
event(ClosingSilverwareDrawer, t_i) \wedge 
\\\nonumber event(OpeningGlasswareCabinet, t_{i+1}) 
\\\nonumber \rightarrow currentActivity(WashDishes, t_i).
\end{eqnarray}
However, the derivation of both activities as instances of \emph{currentActivity} at the same 
time instant $t_i$ would violate formula~(2), making the model inconsistent. 
\end{example}

The statistical-relational approach~\cite{Getoor2007} has recently offered significant advances towards integrating rich expressiveness and uncertainty in one unified framework. MLN is an especially appealing example of this approach. 
The main idea of Markov logic~\cite{Domingos2004} is to allow FOL formulae to be ``softened''.
The validity of a soft formula is evaluated according to the probability of being true with 
respect to a set of axioms describing reality. 
Each soft formula is associated to a \emph{weight} that represents the confidence on 
the validity of the formula.
Weights are generally learned from a training set of
observations.

Formally, a MLN is a pair of two sets~$(\mathcal{F}^S,\mathcal{F}^H)$. Given the signature $S=(O, P, C)$ with $O$ a finite set of typed observable predicate symbols, $P$ a finite set of typed hidden predicate symbols, and $C$ a finite set of typed constants, the set of soft formulae $\mathcal{F}^S$ is a set of $l$ pairs $\{(F_i, w_i)\}, i=1,\ldots,l$ with each $F_i$ being a function-free FOL formula built using predicates from $O \cup P$ and each $w_i \in \mathbb{R}$ a real-valued weight associated with formula $F_i$. The set of hard formulae $\mathcal{F}^H$ is a set of $j$ function-free FOL formulae $\{F_i\}, i=1,\ldots,j$. 
The main inference task of MLN reasoning is to determine the most probable set
of axioms representing reality that can be inferred based 
on the defined formulae and a set of observations (facts).
Intuitively, formulae with higher weight will have higher influence in 
deriving these axioms.

\begin{example}
Referring to Example~\ref{ex:act}, MLN can solve the described problem by assigning 
formula~(1) to the set of soft formulae and formula~(2) to the set of hard formulae. 
The actual weights for the instantiations of formula~(1) 
are estimated through supervised learning on a training set. 
Suppose that the estimated weight of formula (3) is higher than the one of 
formula (4) since, in the training set, the sensor event sequence $\overline{s}$  
is more frequently observed during ``set the table'' than during ``wash dishes''. 
In that case, if $\overline{s}$ is observed, the MLN reasoner would infer that the most 
probable \emph{currentActivity} at $t_i$ is ``set the table''. 
The MLN reasoner cannot derive both activities, since this would violate the hard 
formula~(2). 
Of course, in order to effectively recognize the current activity, we need to use 
multiple formulae involving complex sequences of sensor events.
\end{example}

\subsubsection{Detecting activity boundaries through MLN}
\label{subsubsec:detection}
%
In our model, the observable predicates correspond to the sensor events. 
Predicate $nextEvent(t_i,t_{i+1})$ indicates that the sensor event occurred at $t_i$ and the one occurred at $t_{i+1}$ are consecutive; i.e., the former occurred before the latter, and no other sensor event occurred between them. 
As explained in Section~\ref{subsec:mln}, in our architecture we ensure that no more than one sensor event can occur at a given time instant.
The hidden predicates correspond to the activity boundaries: $startActivity(a,t)$ indicates that activity $a$ begins at time $t$ and $endActivity(a,t)$ indicates that activity $a$ ends at time $t$. 
The approach used for boundary recognition, initially proposed in~\cite{Helaoui2011}, is to write appropriate soft formulae to create a correlation between windows of $n$ consecutive sensor events and start/end of activities. For example, in the case of $n=1$ the following soft formulae can be used:
\begin{itemize}
\item $event($\textit{+}$e,t) \rightarrow startActivity($\textit{+}$a,t)$
\item $event($\textit{+}$e,t) \rightarrow endActivity($\textit{+}$a,t)$
\end{itemize}
Note that \textit{+} symbol before a variable means that a weight is learned for each grounding of that variable.
If we choose $n=2$ the following soft formulae can be used:
\begin{itemize}
\item $event($\textit{+}$e_1,t_1) \land event($\textit{+}$e_2,t_2)  \land \ nextEvent(t_1,t_2)  \rightarrow startActivity($\textit{+}$a,t_1)$
\item $event($\textit{+}$e_1,t_1) \land event($\textit{+}$e_2,t_2) \land \ nextEvent(t_1,t_2)  \rightarrow endActivity($\textit{+}$a,t_2)$
\end{itemize}
For each couple of consecutive sensor events, the first one of the above formulae correlates the first event with the start of an activity; the second formula correlates the second event with an activity end.
In general, the most effective value of $n$ depends on the characteristics of the pervasive system and on the considered activities. In this work, we experimentally choose the optimal value of $n$ (see Section~\ref{subsec:expres}). 

In addition to the soft formulae, we use hard formulae to express common sense knowledge about activity boundaries. In particular, in order to specify that it is impossible that an activity starts and ends at the same time, the following hard formulae are declared: 
\begin{itemize}
\item $startActivity(a,t) \rightarrow \neg endActivity(a,t)$
\item $endActivity(a,t)  \rightarrow \neg startActivity(a,t)$
\end{itemize}

Based on the defined soft formulae, it is possible that the MLN reasoner detects multiple starts for the same activity instance. In order to avoid this issue, we declare the following hard formulae to state that each activity instance cannot start more than once:
\begin{itemize}
\item $currentActivity(a,t_{s}, t) \wedge t \neq t_{s} \\ \rightarrow \neg startActivity(a,t)$,
\end{itemize}
where $currentActivity(a, t_{s}, t)$ states that the instance of activity $a$ that started at 
time $t_s$ is still being performed at time $t$. 
For the sake of this work we do not consider interleaved activities; hence, the hard formulae 
below express that a started activity is carried out until its end:
\begin{itemize}
\item $startActivity(a,t_s) \rightarrow currentActivity(a,t_{s}, t_s)$
\item $currentActivity(a, t_s, t_1) \wedge nextEvent(t_1, t_2) \wedge \\ \neg endActivity(a, t_2)
\rightarrow currentActivity(a, t_s, t_2)$ 
\end{itemize}

Note that in some cases the MLN reasoner may not detect the end of a started activity. This may happen when the patient does not complete that activity at all due to some abnormal behavior; for instance, if the patient sets up the table but forgets to have meal, we consider the activity \emph{having dinner} incomplete. In that case, we post-process the MLN results and we consider the activity ended after a maximum time threshold has expired since its start.

The weights of the soft formulae are learned using a training set of sensor events acquired during the execution of the considered activities. Soft formulae with learned weights and hard formulae are then used to compute \emph{MAP} inference on new data coming from the \textsc{semantic integration layer}.


\subsection{Modeling abnormal behaviors}
\label{subsec:mab}
%
%
As anticipated, our method relies on medical models of abnormal behaviors that may indicate the onset of MCI. 
In order to acquire those models, we collaborated with cognitive neuroscience experts from the Institute Fatebenefratelli\footnote{IRCCS (Research and Care Institute) St John of God Clinical
Research Centre, Brescia -- \url{http://www.irccs-fatebenefratelli.it}}, Lombardy --a leading center in the field of mental health research and research on neurodegenerative disorders-- within the SECURE\footnote{SECURE: Intelligent System for Early Diagnosis and Follow-up at Home, \url{http://secure.ewlab.di.unimi.it/}} research project funded by Lombardy region and MIUR Italian ministry. 
%
Those anomalies have been selected during different project meetings among the technical and medical partners of the SECURE project, based on the medical practice and relevant literature~\cite{scales}. For the sake of this work, we have considered anomalies related to food preparation, food consumption, and compliance to medical prescriptions. The anomalies are defined in natural language by the clinicians; e.g., ``an anomaly occurs when the patient prepares a meal but forgets to consume it''.

\begin{table*}[t!]
\caption{Examples of rules modeling abnormal behaviors}
\label{tbl:rules-examples}
\centering
\begin{tabular}{|c|p{9cm}|p{5cm}|}
\hline
 No. & Rule & Anomaly type \\
\hline
\multirow{2}{*}{1} & $anomaly(cr, fridge, T_1) \leftarrow action(return, RF, S, T_1) \wedge action(close, door, S, T_2) \wedge RefFood(RF) \wedge NonRefStorage(S) \wedge (T_1 < T_2).$ &
 Critical replacement: the patient has placed a food item that needs refrigeration inside a non-refrigerated cabinet.\\
\hline
 \multirow{2}{*}{2} & $anomaly(nca, prepBF, T_1) \leftarrow  startActivity(prepBreakfast, T_1) \wedge endActivity(prepBreakfast, T_2) \wedge ((T_2 - T_1) > 45$minutes$).$ 
 & Non-critical anomaly: the patient spent too much time to prepare breakfast.\\
\hline
 \multirow{2}{*}{3} & $anomaly(co, medicine, null) \leftarrow  prescribed(M, T_1, T_2) \wedge not((action(retrieve, M, C, T) \wedge MedCabinet(C) \wedge (T_1 < T < T_2).$ 
 & Critical omission: the patient has not retrieved a prescribed medicine in due time.\\
\hline
\multirow{2}{*}{4} & $anomaly(wa, medicine, null) \leftarrow  not(prescribed(M, T_1, T_2)) \wedge action(retrieve, M, C, T) \wedge MedCabinet(C) \wedge Medicine(M).$ 
 & Wrong activity: the patient has taken a medicine that was not prescribed.\\
\hline
\end{tabular}
\end{table*}

%
In our model, each IADL performed by a patient consists of a sequence of simple actions, which we call ``steps''. For instance, a patient could perform the IADL ``taking medicines'' by executing this sequence of steps: open the medicine repository, retrieve the medicine box, return the medicine box, close the medicine repository.
In order to facilitate their analysis, we classified anomalies in the following categories:
\begin{itemize}
\item Non-critical anomaly. 
An anomaly is considered as non-critical when the patient skips a relevant step while performing a IADL, or spends too much time to perform the activity, but still he is able to complete the activity correctly. For instance, we consider a non-critical anomaly to occur when the patient forgets to close a repository after taking something from it. Non-critical anomalies can be considered as minor indicators of possible cognitive disorders, only if they occur more frequently than expected, or if their temporal trend significantly increases with time.
\item Critical anomaly. 
A critical anomaly occurs when the patient skips some necessary step while performing an activity, forgets to execute a required activity, or performs the activity more times than expected. Critical anomalies are stronger indicators of possible MCI onset. 
These anomalies are further divided into four categories: 
\begin{itemize}
\item Omission: 
there are some steps in each IADL which are necessary and it is mandatory for the patient to perform them in order to complete the activity correctly: a critical omission occurs when the patient skips any of such steps.  For instance, a critical omission related to the activity ``taking medicines'' is: ``the patient forgets to retrieve the prescribed medicine during the prescribed time'.
%
\item Replacement: this anomaly occurs when, during a IADL, a patient replaces a correct step with a wrong one; for instance, ``the patient has placed the butter inside a non-refrigerated cabinet''.
\item Wrong activity: it occurs when the patient performs an activity that should not be done. For instance, this anomaly occurs when the patient takes a medicine that was not prescribed.
\item Repetition: this anomaly occurs when the patient repeats the same activity more times than expected; for instance, when the patient consumes the morning breakfast twice in a day.
\end{itemize}
\end{itemize}

However, human behaviors are characterized by wide variability; factors such as contextual 
conditions, individual habits and personality traits may determine the execution of 
various anomalies that are not necessarily due to cognitive impairment. This is especially true 
for non-critical anomalies, as leaving repositories open, which may be normally done 
by cognitively healthy people for negligence or hastiness. 

Hence, while the considered anomalies are indicators of possible abnormal behaviors, 
they are not intended to provide an automatic diagnosis of the patient's cognitive status. 
For instance, consider the example of \emph{wrong activity} given above: the fact that
the patient takes a medicine that was not prescribed is critical if he does it
unintentionally (e.g., for a memory disorder). In other cases it may be a normal 
behavior; e.g., if the patient intentionally takes an over-the-counter drug that does 
not interfere with his medical prescriptions. 
Therefore, the frequency of detected anomalies and their temporal trend are 
used as a mean to trigger alarms to the practitioners for further inspecting
the history of detected anomalies and their fine-grained descriptions. 

In order to automatically reason with anomalies, we represent them as rules in propositional logic. 
%
%
%
The anomalies are represented by the predicate \emph{anomaly(Categ, Obj, Time)}. 
\emph{Categ} defines the category of the anomaly.
\emph{Obj} defines the objects or activities involved in the anomaly; for example, in case of a critical omission, the missed medicine may be the object related to that anomaly. 
\emph{Time} defines the time (e.g., day, or exact instant) at which the anomaly has happened. 
%
Table~\ref{tbl:rules-examples} shows the representation of a few anomalies. 
The semantics of \emph{not} is the one of \emph{negation as failure}~\cite{russell}.
Predicate \emph{prescribed($m$,$t_1$,$t_2$)} states that the patient must take medicine 
$m$ from time $t_1$ to time $t_2$ of the current day.
\emph{Medicine(o)} (resp. \emph{Food(o)}) states that object $o$ is a medicine box (resp. 
food item). \emph{Action($a,o,o',t$}) states that the
patient executed action $a$ on objects $o$ and $o'$ at time $t$. 
\emph{Holds($s$,$o$,$t_1$,$t_2$)} states that the status of object $o$ has been ``$s$'' 
from $t_1$ to $t_2$ (for instance, ``the microwave oven has been on from 11:30 to 11:55'').
The Holds predicate allows us to express temporal conditions that are useful in the 
definition of different anomalies. Temporal expressions that we use in our rule-based 
definitions include the interval of time during which an action is performed, the temporal 
distance between two actions, the temporal duration of an activity, the temporal order
among activities.

\subsection{Recognizing abnormal behaviors}
Abnormal behaviors are recognized by a knowledge-based inference engine,
which periodically (e.g., at the end of each day) evaluates the rule-based 
anomaly definitions considering 
the data acquired and inferred during the considered time period:
activity boundaries, actions and sensor events, as well as external 
knowledge including the medical prescriptions of the patient and the classification
of objects in categories. 
Those data are represented by the predicates introduced in 
Section~\ref{subsec:mab}, and added to the propositional 
logic knowledge base.

\begin{example}
%
%
%
%
Consider an elderly person living independently at home. 
Suppose that furniture and devices, including food cabinets and the fridge, 
are equipped with a magnetic sensor to detect the \emph{open}
and \emph{close} actions.
An RFID tag is attached to some food boxes to identify their content (e.g., 
rice, milk, coffee, sugar). RFID readers in the proximity of the cabinets
and fridge are in charge of 
detecting which item has been retrieved or returned. 
Suppose that at 08:05 AM the patient opens the fridge $f$ and retrieves the milk
box $m$ to prepare breakfast. Then after a few minutes he mistakenly puts the milk box 
in the non-refrigerated food cabinet $c$ and closes its door. 
Hence, based on the sensed events, the following axioms are 
automatically added to the knowledge base:
\begin{eqnarray}\nonumber
\textrm{\emph{action(open, door, $f$, $8$:$05$:$00$ AM).}}\\\nonumber
\textrm{\emph{action(retrieve, $m$, $f$, $8$:$05$:$07$ AM).}}\\\nonumber
\textrm{\emph{action(return, $m$, $c$, $8$:$12$:$30$ AM).}}\\\nonumber
\textrm{\emph{action(close, door, $c$, $1$:$12$:$35$ AM).}}\nonumber
\end{eqnarray}
Since the knowledge base contains the axioms \emph{RefFood(m)} (stating that
the milk box contains a food item that must be kept refrigerated) 
and \emph{NonRefStorage(c)} (stating that $c$ is a non-refrigerated storage), 
rule $1$ in Table~\ref{tbl:rules-examples} fires,
recognizing an abnormal behavior.
%
%
\end{example}


\section{Experimental evaluation}
\label{sec:experiments}
In order to evaluate the FABER technique we developed a prototype
of the system in a smart home lab, we acquired a large dataset of both normal and abnormal
behaviors, and we executed experiments to evaluate the effectiveness of the system.

\subsection{Software implementation}
\begin{figure*}[t!]
\centering
\subfigure[Android application running the FABER software]{\label{fig:app}
\includegraphics[width=.17\linewidth]{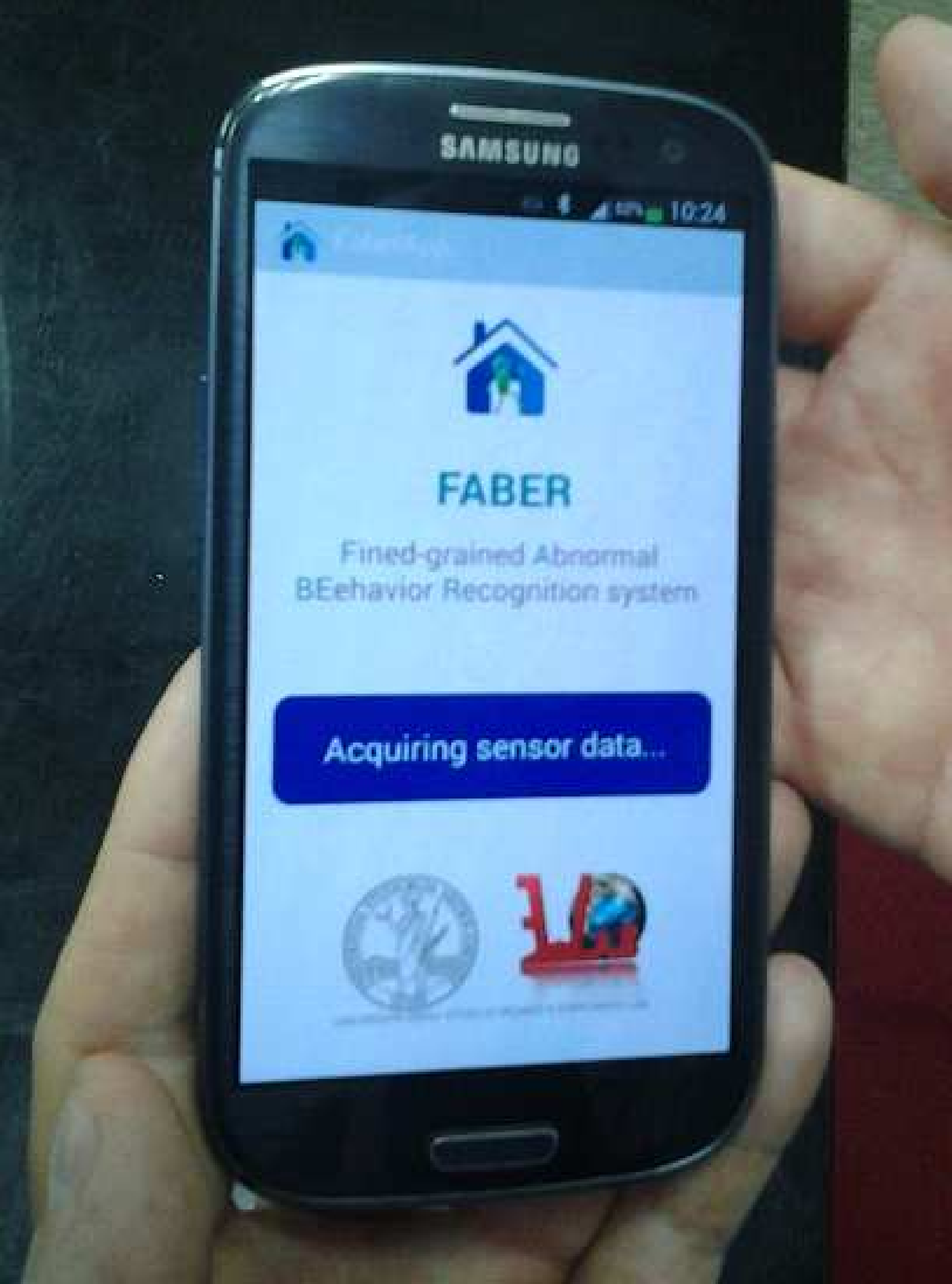}
}
\subfigure[Dashboard interface to inspect the status of the smart home infrastructure]{\label{fig:dash-inf}
\includegraphics[width=.4\linewidth]{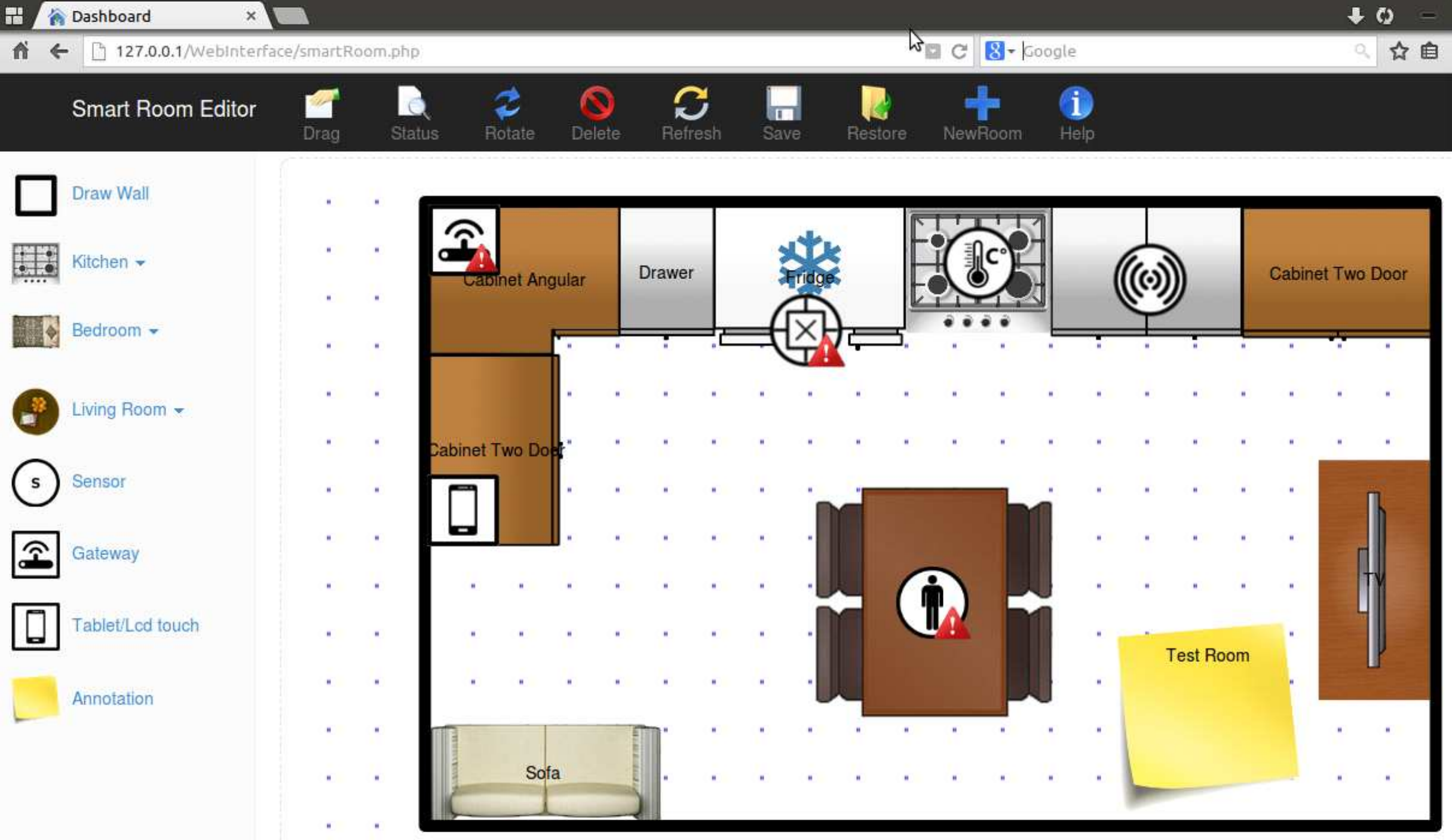}
}
\subfigure[Dashboard interface to view the temporal trend of anomalies]{\label{fig:dash-trend}
\includegraphics[width=.37\linewidth]{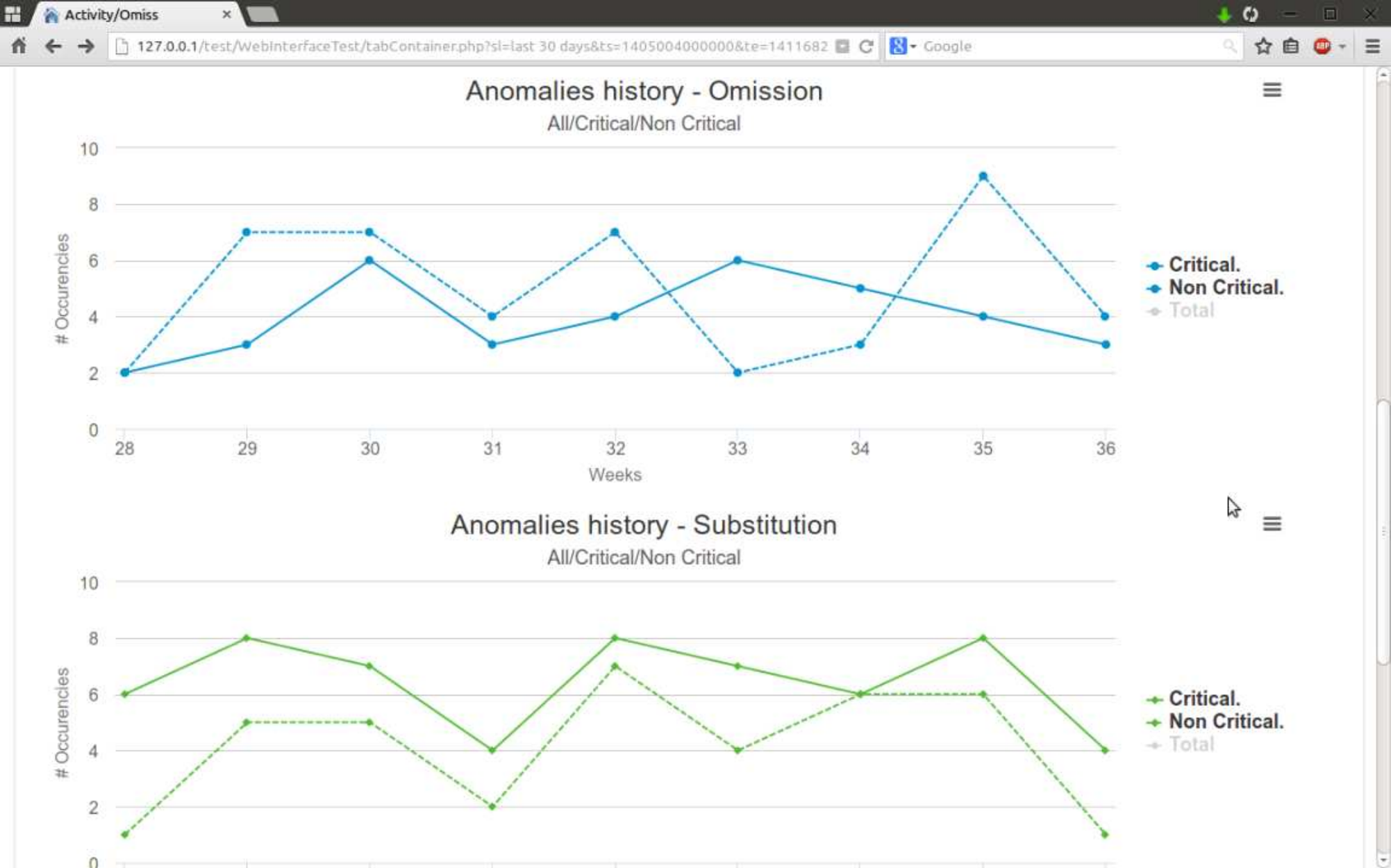}
}
\caption{Android app and Web dashboard of the FABER system.\label{fig:software}}
\end{figure*}
A prototype implementation of the whole system has been developed within the
activities of the SECURE project.
Since the FABER system is intended to run on a mobile device at the patient's
home, the core software modules have been implemented in Java for the Android
platform. Figure~\ref{fig:app} shows the application running the FABER
software. 
In particular, in order to implement the technique for activity boundary
detection we have used the Java libraries of Markov TheBeast~\cite{riedel2008}, 
which allow to solve MAP integer linear programming problems through 
a cutting plane inference meta algorithm.
In order to evaluate the rule-based definitions of anomalies we used the APIs
of TuProlog~\cite{tuprolog}, a lightweight Java implementation of an inference 
engine for the well-known Prolog logic programming language.

Currently, most sensor motes available on the market communicate using the 
ZigBee protocol, and there is no standard interface for that protocol on
most Android devices. Hence, we use a gateway installed in the
smart home to receive ZigBee messages from sensors and forward them via 
Bluetooth to the Android device.
Sensor motes have been programmed in the C++ language to communicate new
data to the gateway at the occurrence of each event of interest. For instance, the pressure
sensor attached to the kitchen chair seat communicates the measured pressure 
when it exceeds or falls behind some given thresholds, to detect when the patient 
stands up or sits down on the chair. Such thresholds have been determined 
empirically. 
The sensor event message includes the timestamp of the sensor reading, the
sensor ID and the detected value.
In the current implementation, we use the Libelium Meshlium gateway, 
which runs a Linux OS. A C++ application running on
the gateway is in charge of: receiving data from sensors, assigning the
unique timestamps, locally storing the data 
in a PostgreSQL database, and periodically communicating the data to the 
Android application. At the end of each day, the Android app runs the FABER 
algorithms for activity boundary detection and anomaly recognition, and 
communicates the results through the Internet to the backend of the hospital 
center, where data are stored.

We have also developed a Web-based dashboard,
shown in Figures~\ref{fig:dash-inf} and~\ref{fig:dash-trend}, 
to allow technicians to inspect the
status of the smart home infrastructure in order to identify possible issues 
(sensor failures, sensor battery exhaustion\ldots), and to allow practitioners
to analyze the history and trends of IADLs and abnormal behaviors.

\subsection{Smart room environment}
%
We have instrumented a room in a smart home lab with different
kinds of sensors to detect low-level actions and events.

\begin{table}[!ht] 
\caption{Monitored household items}
\label{table:sensors}
\centering
\begin{tabularx}{.9\columnwidth}{|X|l|}
\hline
\textbf{Monitored items} & \textbf{Related sensors} \\
\hline\hline
Medicines boxes, 
Food items containers & {RFID readers} \\
\hline
Medicines cabinet, 
Fridge, 
Non-refrigerated food cabinet, 
Cooking pan cabinet, 
Silverware drawer & {Magnetic sensors} \\
\hline
Stove & Temperature sensor \\
\hline 
Kitchen table & {Presence sensor} \\
\hline
Kitchen chair & Pressure sensor \\ 
\hline 
\end{tabularx} 
\end{table}

\begin{figure*}[t!]
\centering
\subfigure[Magnetic sensor attached to a drawer]{\label{fig:drawer}
\includegraphics[height=.21\linewidth]{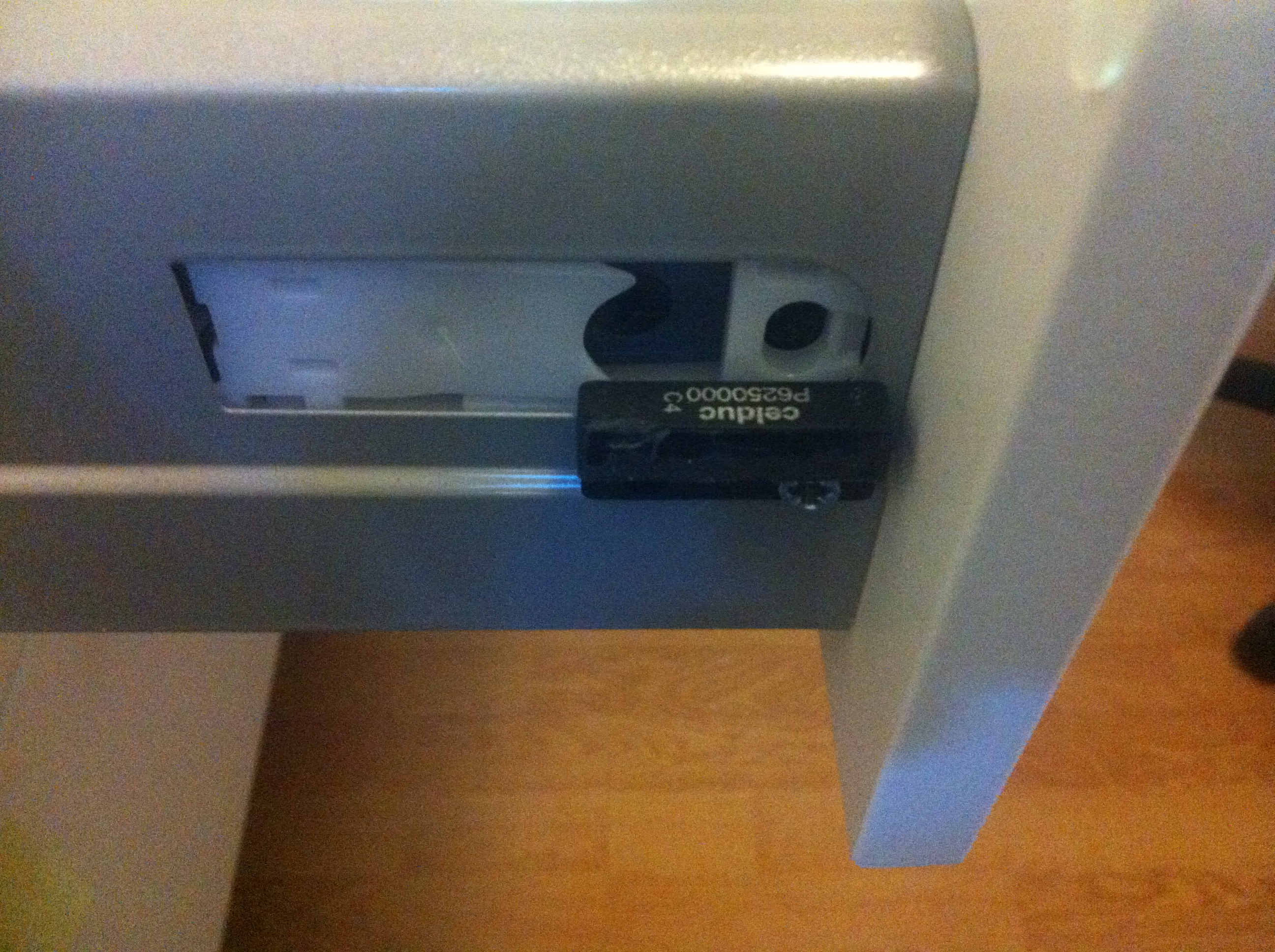}
}
\subfigure[Presence sensor above the kitchen table]{\label{fig:presence}
\includegraphics[width=.28\linewidth]{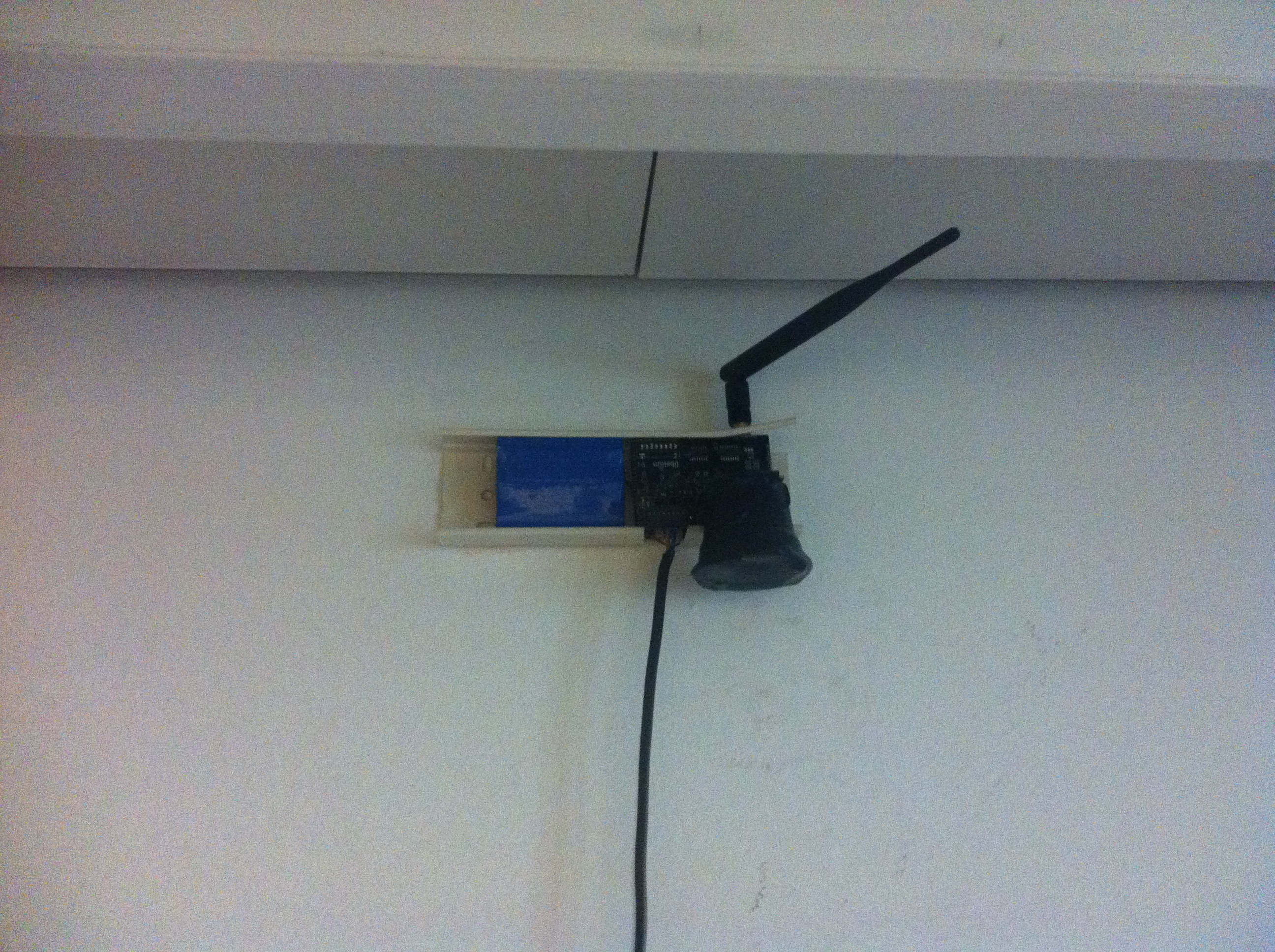}
}
\subfigure[RFID reader for medicine boxes and food items]{\label{fig:rfid}
\includegraphics[width=.35\linewidth]{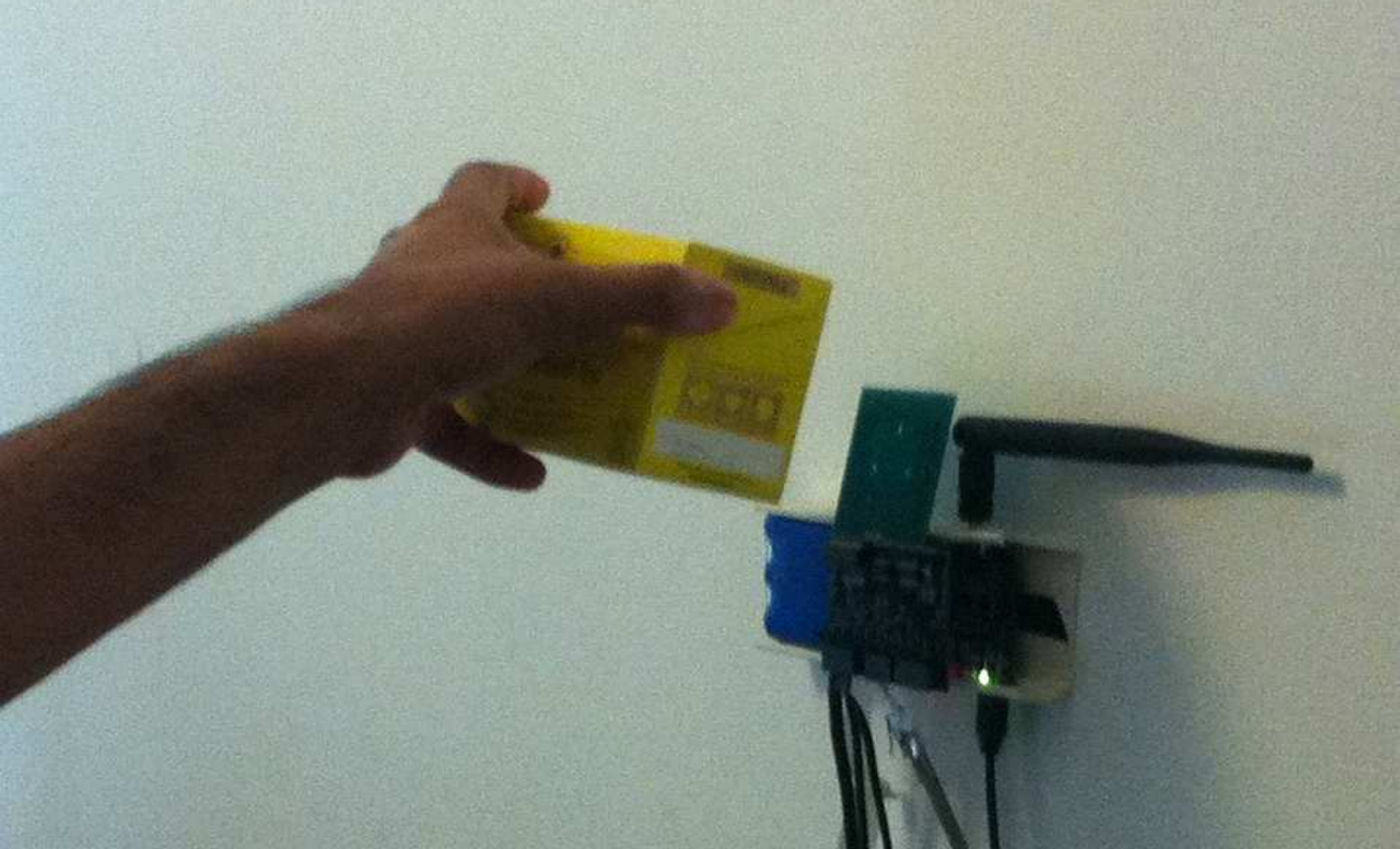}
}
\caption{Some sensors used in the smart home lab.\label{fig:sensors}}
\end{figure*}

Table \ref{table:sensors} shows the different kinds of sensors that have been deployed on various household items. 
Some of the used sensors are illustrated in Figure~\ref{fig:sensors}. 
\emph{RFID} tags are attached to medicines and food packages. In the current prototype we assume that whenever a patient retrieves or returns an item with an attached tag, he passes the item's tag near the \emph{RFID} reader, in order to let the system identify the object. Of course, this method is inconvenient, especially for elderly people; in the future this method will be replaced exploiting more convenient tracking technologies. 
Magnetic sensors are used to monitor opening and closing of various repositories (e.g.; fridge, medicine cabinet\dots). A temperature sensor is used to detect if the patient is using the stove. A presence sensor is used to monitor the presence of the patient in the proximity of the dining table. A pressure sensor on the seat is used to understand if the patient is sitting at the kitchen chair.


%

\subsection{Dataset acquisition}
We have acquired a dataset of IADLs and anomalies, asking to voluntary actors to reproduce the daily routine of $21$ patients in our smart home lab. Executed IADLs and anomalies have been carefully designed in collaboration with neuroscience experts to realistically reproduce the behavior of two groups of elderly persons: $7$ healthy seniors (group 1), and $14$ elderly people with early symptoms of MCI (group 2). We assume that individuals of both groups live alone in their respective homes. During his one-day routine, each individual in group 1 does not execute any critical anomaly, but may execute a few non-critical ones. Individuals in group 1 are mainly used to evaluate the number of false positives produced by our anomaly recognition method. Group 2 individuals may perform several non-critical and critical anomalies during the day. 

During the execution of the activities and anomalies, we have acquired the timestamped data coming from the sensors deployed in the smart home and manually annotated the dataset with the start- and end-time of activities and anomalies. 
The following IADLs have been selected to validate our method: 
\begin{itemize}
\item Preparing food: the patient has to prepare the daily meals (breakfast, lunch, dinner) at appropriate times.
\item Consuming meal: when the patient prepares a meal, he has to consume it within a reasonabile time period.
\item Taking medicines: the patient has to take the prescribed medicines in the due time. We assume that no smart dispenser is used; instead, we assume that the patient keeps all the medicines in a dedicated cabinet.
\end{itemize}
We have considered the following anomalies:
\begin{itemize}
\item Non-critical anomalies. They happen when the individual: (NC1) forgot a repository open; (NC2) did not return a medicine to its cabinet; (NC3) retrieved a food item which must be cooked, but did not use the stove burner; (NC4) does not prepare a meal.
\item Critical anomalies. They happen when the individual: (C1) did not retrieve a prescribed medicine in the due time; (C2) took a medicine that was not prescribed; (C3) took a prescribed medicine in the due time but multiple times, resulting in inappropriate dosage; (C4) did not turn off the stove burner after finishing to prepare a hot meal; (C5) did not take the silverware before consuming meal; (C6) did not consume the meal after having prepared it; (C7) turned the stove burner on but did not take any cooking pan.
\end{itemize}

Totally, our dataset contains $21$ days of IADLs and anomalies. Group 1 individuals did 7 non-critical and 0 critical anomalies; group 2 individuals did 29 non-critical and 28 critical anomalies.

\subsection{Experimental setup}
We experimentally evaluated the FABER method using
a $21$-fold cross-validation process. The dataset was partitioned into 
$21$ portions, each corresponding to the data acquired during the one-day
activities of a different individual. For each experiment we used $20$   
portions as training set and the remaining one as test set. This process was
iterated $21$ times, using each partition exactly once as test set. 
The prediction's quality was evaluated in terms of the standard measures of precision,
recall and $F_1$ score (the latter is the harmonic mean of precision and recall). 

Since the anomaly recognition technique relies on detected activity boundaries,
we first needed to experimentally choose the value of parameter $n$, 
corresponding to the length of the temporal sequence of sensor events to be used by 
our MLN-based method.
Then, we applied both the activity boundary detection method using the chosen $n$ value, 
and the anomaly recognition technique to evaluate the effectiveness of the whole technique.

\subsection{Results}
\label{subsec:expres}
%
%

\subsubsection{Choice of the MLN model}
As explained in Section~\ref{subsubsec:detection}, activity boundary detection relies on a trained \emph{MLN} that analyzes temporal sequences of $n$ consecutive sensor events. 
The optimal choice of $n$ depends on many factors, including the monitored activities, the characteristic of the environment and the available sensors. Hence, we performed a $21$-fold cross-validation process to experimentally choose the $n$ value.

\begin{figure}[t!]
  \centering
     \includegraphics[width=.75\columnwidth]{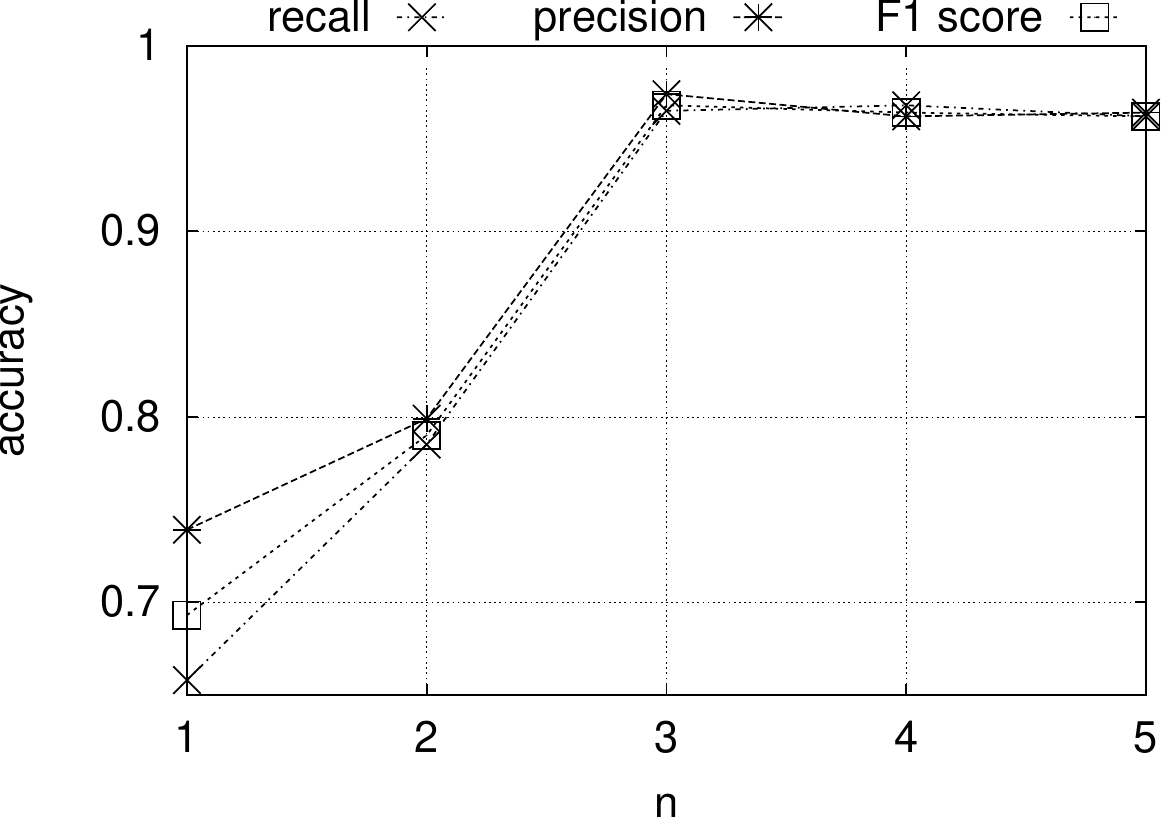}
  \caption{Accuracy of the activity boundary detection technique; $n$ is the length of the considered temporal sequence of sensor events}
  \label{fig:plot-cv}
\end{figure}

Results are shown in Figure~\ref{fig:plot-cv}. With $n=1$, the recognition performance
was quite low; this indicates that it is impossible to accurately detect the start/end
times of activities based on one-sensor data only. The reason is that the same sensor event 
may be fired by the execution of different IADLs, or by the execution of both the
start and the end of a given activity. We can observe that the value of recall was lower
than the one of precision. This means that the MLN-based method, which tries to maximize the
probability of the predicted boundaries, 
failed to recognize several boundaries due to the
lack of sufficient information to support their prediction. Hence, it obtained a large
number of false negatives but a lower number of false positives.

The overall accuracy improved using $n=2$. This result indicates that longer
temporal sequences of sensor events are stronger indicators of IADLs start-
and end-times. Indeed, with $n=3$ the system achieved good levels of detection,
with precision, recall and $F_1$ score larger than $0.96$. Using values of $n$ larger
than $3$ did not provide advantages in terms of accuracy, while complicating the MLN model. 
Hence, in the following experiments we set the value of $n$ to $3$ and
we used the following soft formulae:
\begin{itemize}
\item $event($\textit{+}$e_1,t_1) \land event($\textit{+}$e_2,t_2)  \land \ nextEvent(t_1,t_2)  \rightarrow \\startActivity($\textit{+}$a,t_1)$
\item $event($\textit{+}$e_1,t_1) \land event($\textit{+}$e_2,t_2)  \land \ nextEvent(t_1,t_2)  \rightarrow \\startActivity($\textit{+}$a,t_2)$
\item $event($\textit{+}$e_1,t_1) \land event($\textit{+}$e_2,t_2) \land \ nextEvent(t_1,t_2)  \rightarrow \\endActivity($\textit{+}$a,t_1)$
\item $event($\textit{+}$e_1,t_1) \land event($\textit{+}$e_2,t_2) \land \ nextEvent(t_1,t_2)  \rightarrow \\endActivity($\textit{+}$a,t_2)$
\item $event($\textit{+}$e_1,t_1) \land event($\textit{+}$e_2,t_2) \land event($\textit{+}$e_3,t_3) \land \ nextEvent(t_1,t_2) \land \ nextEvent(t_2,t_3)  \rightarrow \\startActivity($\textit{+}$a,t_2)$
\item $event($\textit{+}$e_1,t_1) \land event($\textit{+}$e_2,t_2) \land event($\textit{+}$e_3,t_3) \land \ nextEvent(t_1,t_2) \land \ nextEvent(t_2,t_3)  \rightarrow \\endActivity($\textit{+}$a,t_2)$
\end{itemize}

\subsubsection{Anomaly recognition}
\begin{table}[th!]
\caption{Results of abnormal behavior recognition}
\label{tbl:statistics}
\centering
\begin{tabular}{|l|l;{2pt/2pt}l;{2pt/2pt}l|d{0};{2pt/2pt}l;{2pt/2pt}l|}
\hline
\multirow{2}{*}{ANOMALY} & \multicolumn{ 3}{c|}{{GROUP 1}} & \multicolumn{ 3}{c|}{{GROUP 2}} \\
 & {TP} & {FP} & {FN} & \textrm{TP} & {FP} & {FN} \\
\hline\hline
NC1: Repository left open & 5 & 0 & 2 & 14 & 0 & 0 \\
\hline
NC2: Medicine not returned & 0 & 0 & 0 & 4 & 0 & 0 \\
\hline
NC3: Food item not cooked & 0 & 0 & 0 & 2 & 0 & 0 \\
\hline
NC4: Meal not prepared & 0 & 2 & 0 & 0 & 1 & 0 \\
\hline
C1: Missed a prescr. medicine & 0 & 2 & 0 & 10 & 0 & 0 \\
\hline
C2: Took a wrong medicine & 0 & 0 & 0 & 7 & 0 & 0 \\
\hline
C3: Repeated medicine intake & 0 & 0 & 0 & 3 & 0 & 0 \\
\hline
C4: Stove burner left on & 0 & 0 & 0 & 0 & 0 & 0 \\
\hline
C5: Had meal with no silverware & 0 & 0 & 0 & 7 & 0 & 0 \\
\hline
C6: Prepared meal not consumed & 0 & 0 & 0 & 1 & 1 & 0 \\
\hline
C7: Burner turned on by mistake & 0 & 0 & 0 & 8 & 0 & 0 \\
\hline\hline
\multirow{1}{*}{TOTAL} & 5 & 4 & 2 & 48 & 2 & 0 \\
\hline
\end{tabular}
\end{table}
\begin{table}[th!]
\caption{Precision, recall and F1 score}
\label{tbl:acc}
\centering
\begin{tabular}{|l|d{3}|d{3}|d{3}|}
\hline 
ANOMALY TYPE  & \multicolumn{1}{c|}{PRECISION} & \multicolumn{1}{c|}{RECALL} & \multicolumn{1}{c|}{F1 SCORE} \\
\hline \hline
Non-critical & 0.893 & 0.926 & 0.909 \\
\hline
Critical & 0.923 & 1 & 0.96 \\
\hline \hline 
TOTAL & 0.898 & 0.964 & 0.93 \\
\hline
\end{tabular}  
\end{table}
Results of anomaly recognition are reported in Table~\ref{tbl:statistics}. 
Each row of the table corresponds to a specific anomaly considered in our
experiments. The TP column reports the number of true positives for that
anomaly; i.e., the number of actual occurrences of that anomaly that were 
recognized by FABER.
FP reports the number of false positives; i.e., the number of anomalies
reported by FABER that did not actually occur.
FN reports the number of actual occurrences of that anomaly that were not
recognized by FABER.

As anticipated, group 1 individuals performed a few non-critical anomalies
(NC) and no critical anomaly (C). The system correctly recognized $5$ NCs
out of $7$. Looking closely at the data, we discovered that in two cases the
system did not detect NC1 (repository left open) due to a failure of the
magnetic sensor, which did not communicate the opening of the medicine 
drawer. This kind of issue can be addressed by introducing redundancy in
the sensing infrastructure. 
During the 7 days activities of group 1 individuals, the system
did $4$ false positives. Two of them regarded NC4 (meal not prepared), while
the other two regarded C1 (missed a prescribed medicine). These errors were 
due to mispredictions of the activity boundary detection technique, which
in two cases did not recognize the occurrence of activity ``preparing meal''
and in two cases did not recognize the occurrence of ``taking medicine''. 
Hence, the FP rate could be reduced by using more and/or better sensors,
as well as more effective activity recognition methods, to
improve the performance of the activity boundary detection technique.

Group 2 individuals performed a larger number of NCs and several Cs.
For this group, FABER correctly recognized all the occurrence of both
critical and non-critical anomalies; i.e., no false negatives happened. 
During the 14-days activities of that group, the system reported only 2 false 
positives: one was related to NC4 (meal not prepared) and the other one 
to C6 (prepared meal not consumed). Even in these cases, false positives 
were due to mispredictions of the activity boundary detection technique. 

Overall, the system produced 6 false positives during the 21-days activities
of the two groups. We claim that the number of false positives is compliant 
with the requirements of clinicians, especially considering that the 
individuals totally performed more than $150$ instances of activities 
during the experimentation.
Table~\ref{tbl:acc} reports the results in terms of precision, recall
and F1 score. The achieved precision was close to $0.9$. This relatively low 
value was mainly due to missed activity boundary detection by the MLN-based 
technique. However, the precision of critical anomaly
recognition is significantly higher than the one of non-critical anomaly
recognition. 
When activity boundaries were correctly recognized, in most cases FABER 
recognized the occurred anomalies, achieving an overall recall
larger than $0.96$.
A preliminary clinicians' assessment of the FABER system can be found
in~\cite{smarte15}.

\section{Conclusions and future work}
\label{sec:conclusions}
In this paper we addressed the challenging issue of unobtrusively
recognizing abnormal behaviors exhibited by elderly persons at
home. We have proposed the FABER hybrid technique to recognize anomalies 
at a fine-grained level, based on the integration of supervised learning 
and symbolical reasoning, and on sensor data acquired from the smart-home
infrastructure. Differently from existing approaches, our method provides
detailed information about the detected anomalies, which can be
exploited by practitioners for early detection of MCI. 
We designed the models of anomalies collaborating with cognitive 
neuroscience experts, and we implemented a 
prototype of FABER in a smart home lab. Experiments with a 
dataset of activities and anomalies show that FABER achieves high recall 
while generating a small number of false positives.

The achieved results are promising, but we plan to improve this work
in several directions. 
Since both activity recognition and sensor data acquisition are prone to 
inaccuracy, a first improvement may be extending our technique to more 
extensively support reasoning with uncertainty. Currently we use MAP 
inference to compute the most probable activity boundaries; hence, the 
predicted boundaries are not associated to a confidence level. We will 
investigate different MLN inference methods (e.g., marginal inferencing) 
to compute probabilistic activity boundaries. More importantly, our 
current anomaly recognition method is based on non-probabilistic rules 
that strictly determine the detection of an abnormal behavior based on a 
user-defined set of observations. We consider extending this rigid 
system with probabilistic reasoning, possibly by means
of a probabilistic logic. 
Other future work includes addressing the case of multi-inhabitants,
concurrent and interleaved activities. 
Finally, we are working closely with clinicians to extend the set of 
significant anomalies to be monitored and we are already conducting 
experiments in the patients' homes.


\section*{Acknowledgments}
This work has been partially supported by the project 
``SECURE: Intelligent System for Early Diagnosis and Follow-up 
at Home'', funded by a grant of Lombardy Region and Italian 
Ministry of Education, University and Research.

\end{document}